\documentclass[prl, aps, amsfonts, twocolumn, showpacs]{revtex4}
\usepackage{graphicx,subfigure}
\usepackage{amsmath}

\renewcommand{\d}{\partial}

\newcommand{\lp}{\bigg(}
\newcommand{\rp}{\bigg)}

\newcommand{\pf}[2]{\lp\frac{#1}{#2}\rp}

\newcommand{\mean}[1]{\langle#1\rangle}

\newcommand{\sinc}{{\rm sinc} }

\newcommand{\be}{\begin{equation}}
\newcommand{\ee}{\end{equation}}

\begin{document}

\title{General Relation between Entanglement and Fluctuations in One Dimension}
\author{H.~Francis~Song}
\author{Stephan~Rachel}
\author{Karyn~Le~Hur}
\affiliation{Department of Physics, Yale University, New Haven, CT 06520}

\begin{abstract}
In one dimension very general results from conformal field theory and exact calculations for certain quantum spin systems have established universal scaling properties of the entanglement entropy between two parts of a critical system. Using both analytical and numerical methods, we show that if particle number or spin is conserved, fluctuations in a subsystem obey identical scaling as a function of subsystem size, suggesting that fluctuations are a useful quantity for determining the scaling of entanglement, especially in higher dimensions. We investigate the effects of boundaries and subleading corrections for critical spin and bosonic chains.

\end{abstract}


\maketitle

Entanglement entropy, which measures nonlocal correlations in a quantum system, plays an important role in such diverse areas as the study of black holes \cite{wilczek} and quantum computation \cite{nielsen}. More recently, much attention has been focused on entanglement in  condensed matter systems \cite{rmp} and in particular on the role of entanglement in quantum phase transitions at zero temperature \cite{nature}. A significant discovery arising from this investigation has been the universal scaling of entanglement entropy in one-dimensional quantum critical systems described by conformal field theory (CFT) \cite{vidal,calabrese,wilczek,disorder}. Despite these advances, the experimental relevance of these theories has remained unclear; the same feature that makes this quantity so universal---mainly, the fact that entanglement entropy is defined without reference to the observables of a system---has precluded its measurement in real quantum many-body systems. Recently, however, it was shown \cite{levitov} that for the special case of free fermions the entanglement entropy can be related exactly to the full set of cumulants of the charge fluctuations, suggesting that entanglement entropy could be accessed through the fluctuations.

In this Letter we propose that the fluctuations of a conserved charge is an interesting quantity to study in relation to entanglement entropy beyond the free-fermion case. In particular, we show that in one-dimensional critical systems with particle number or spin conservation, the variance of the fluctuations in a subsystem (henceforth simply ``fluctuations'') behaves very similarly to the entanglement entropy \emph{even when there are interactions}. Fluctuations, like entanglement entropy, diverge logarithmically as a function of subsystem size in conformally invariant systems with a globally conserved charge.

The entanglement entropy of a subsystem $A$ of size $x$ embedded in a larger system of size $L$ is given by the von Neumann entropy $\mathcal{S}(x,L)=-\text{Tr }\hat{\rho}_A\ln\hat{\rho}_A$ of the reduced density matrix $\hat{\rho}_A$ for subsystem $A$. For a critical system described by a CFT with central charge $c$, the entanglement entropy at zero temperature for $L\rightarrow\infty$ was shown to have the universal behavior \cite{calabrese}
\be
	\mathcal{S}(x) = \frac{c}{3}\ln x + s_1\label{eq:Skorepin},
\ee where $s_1$ is a non-universal constant. For later comparison it is useful to note that this was achieved by computing the quantity $\text{Tr }\hat{\rho}_A^n$ and differentiating with respect to $n$ at $n=1$. The versatility of this formalism lay in the fact that $\text{Tr }\hat{\rho}_A^n$ transforms simply under conformal mappings, allowing $\mathcal{S}(x,L)$ to be computed for finite $L$, finite temperature, and different boundary conditions.

Now for the same setup, consider the number fluctuations in subsystem $A$,
\be
	\mathcal{F}_A = \mean{(\hat{N}_A-\mean{\hat{N}_A})^2}\label{eq:Fdef},
\ee where $\hat{N}_A$ counts the number of particles in subsystem $A$. For spins we replace $\hat{N}$ by $\hat{S}^z$. We expect that the fluctuations will behave similarly to the entanglement entropy, since for a pure state $\mathcal{F}_A=\mathcal{F}_B$ where $B$ is the remainder of the system. As noted for entanglement entropy, this symmetry implies that generically, the fluctuations reside mainly on the boundary between the two subsystems, leading to an area law with possible logarithmic corrections \cite{arealaw}. Indeed, it is easy to show that all even-order cumulants of the particle number satisfy this property, which explains the absence of odd cumulants in the formula of Ref.~\onlinecite{levitov}. Moreover, $\mathcal{F}_A=0$ for separable states, and for valence bond states $\mathcal{F}_A$ coincides (up to a factor 1/4) with both the von Neumann and valence bond entropies \cite{vb}. On the practical side, for most systems in any dimension the fluctuations are easier to compute numerically than the valence bond entanglement entropy, which was introduced partly for its computational convenience relative to the von Neumann entropy.

We first consider Luttinger liquids (LLs), which describe the low-energy properties of many one-dimensional systems \cite{giamarchi}. From LL theory in the limit $L\rightarrow\infty$ we have $\pi^2\mathcal{F}_\text{LL}=\mean{[\phi(x)-\phi(0)]^2}$ where $\phi$ is the ``charge'' field, so that at zero temperature
\be
\pi^2\mathcal{F}_\text{LL}(x)
	= K\ln \frac{x}{a}\label{eq:LL},
\ee with $K$ the Luttinger parameter and $a$ a short-distance cutoff. As for entanglement entropy \cite{calabrese}, the same result with $K\rightarrow K/2$ and $x\rightarrow 2x$ is obtained when there is a boundary, due to the constraint $\phi(0)=$ constant. An interesting confirmation of the LL result comes from the $\nu=1/m$ fractional quantum Hall states \cite{laughlin}, for which $K=\nu$. A detailed calculation \cite{qh} shows that in the time domain, charge fluctuations across a quantum point contact with quantum Hall wires are given by $\pi^2\mathcal{F}(t)=\nu \ln (t/\delta)$ with short-time cutoff $\delta$, as might be expected from Eq.~(\ref{eq:LL}) and Lorentz-invariance.

The logarithmic scaling of fluctuations extends beyond LLs, and holds generally for critical models described by a CFT with a conserved U(1)-current (i.e., fixed total particle number or spin component), which is always described by a massless free boson. As for entanglement entropy these results therefore extend simply to finite size, finite temperature, and different boundary conditions via conformal mapping. To see this, note that if we define the characteristic function
\be
	M_A(\lambda) = \mean{e^{i\lambda(\hat{N}_A-\mean{\hat{N}_A})}}
	= \pf{x}{a}^{-g\lambda^2/(2\pi^2)}
	\label{eq:MA}
\ee which transforms simply under conformal mappings, then $\pi^2\mathcal{F}_A=-\pi^2 M_A''(0)= g\ln(x/a)$. The prefactor $g$ can always be fixed by considering the physical meaning of the charge, but we can give a heuristic argument for its value as follows. At finite temperature $1/\beta$ (we set $\hbar=k_B=v=1$, where $v$ is the effective velocity) the mapping $z\rightarrow z'=(\beta/2\pi)\ln z$ in Eq.~\eqref{eq:MA} gives
\be
	\pi^2\mathcal{F}(x,\beta)=g \ln\lp \frac{\beta}{\pi a}\sinh \frac{\pi x}{\beta} \rp\label{eq:Fbeta}.
\ee For sufficiently large $x\gg \beta$ such that interactions across the boundary can be neglected (which is possible since correlations decay exponentially), we may consider the subsystem $A$ to be a grand canonical ensemble in equilibrium with a bath consisting of the remainder of the system \cite{bell}. This is of course only possible if the total particle number is fixed. Then from standard statistical mechanics one has $\mathcal{F}(x,\beta)\sim \kappa x/\beta$ where $\kappa=\d n/\d\mu$ is the compressibility (susceptibility $\chi=\d m/\d B$ for spins), so that by matching Eq.~\eqref{eq:Fbeta} for $x\gg\beta,a$ we find
\be
	g = \pi v \kappa\label{eq:gfactor}.
\ee We have put in the velocity $v$ for completeness. Note that this is consistent with the LL expression since $K=\pi v\kappa$ \cite{giamarchi}. Eq.~\eqref{eq:gfactor} turns out to be quite general, as we will see later. These results generalize the logarithmic scaling of fluctuations noted for noninteracting fermions \cite{klich} to conformally invariant, interacting systems. Interestingly, to leading order the entanglement entropy and fluctuations both obey logarithmic scaling, and
\be
	\frac{\mathcal{S}(x)}{\pi^2\mathcal{F}(x)} \sim \frac{c}{3\pi v\kappa}, \qquad x\gg a \label{eq:SFratio}.
\ee This generalizes Ref.~\onlinecite{levitov}, with the ratio modified by the central charge and compressibility, for the free bosonic theory where only the second cumulant is non-zero. Of course, there are subleading corrections to both quantities, which we study in detail. In many cases the subleading terms of the entanglement entropy and fluctuations also behave similarly, especially when there are boundaries. Indeed, in the presence of boundaries the logarithmic prefactors for both quantities are half of their periodic values, so that Eq.~\eqref{eq:SFratio} remains unchanged.

In the following we study the detailed behavior of fluctuations in several important models, including systems that are not described by LLs.

\emph{XXZ Model.}---We first consider the spin-1/2 XXZ Hamiltonian
\be
	H_\text{XXZ} = \sum_i (\hat{S}^x_i\hat{S}^x_{i+1} + \hat{S}^y_i\hat{S}^y_{i+1} + \Delta \hat{S}^z_i\hat{S}^z_{i+1})\label{eq:Hxxz}
\ee for $-1<\Delta\leq1$ where the model is gapless (the point $\Delta=-1$ is not conformally invariant). At $\Delta=0$ it reduces to an exactly solvable problem of free fermions. Using the Jordan-Wigner transformation we can use the methods of Ref.~\onlinecite{lieb} to compute the correlation matrix $G_{ij}$, which in the limit $L\rightarrow\infty$ is found to be $G^\text{pbc}_{ij} = \delta_{ij}-\sinc[\pi(i-j)/2]$ for periodic boundary conditions (PBCs) and $G^\text{obc}_{ij} = \delta_{ij}-\sinc[\pi(i-j)/2]+\sinc[\pi(i+j)/2]$ for open boundary conditions (OBCs), where $\sinc\ x=1$ for $x=0$ and $\sin x/x$ otherwise. For exact diagonalization the slightly more complicated finite-$L$ expressions were used. The entanglement entropy was numerically computed as in Ref.~\onlinecite{vidal}, while $\mathcal{F}_\text{XX}(\ell)=\sum_{i,j=1}^\ell [\mean{\hat{S}^z_i\hat{S}^z_j} - \mean{\hat{S}^z_i}\mean{\hat{S}^z_j}]=(\ell-\sum_{i,j=1}^\ell G_{ij}^2)/4$ for a block of $\ell$ sites.

\begin{figure}[tp]
\includegraphics*[width=215pt]{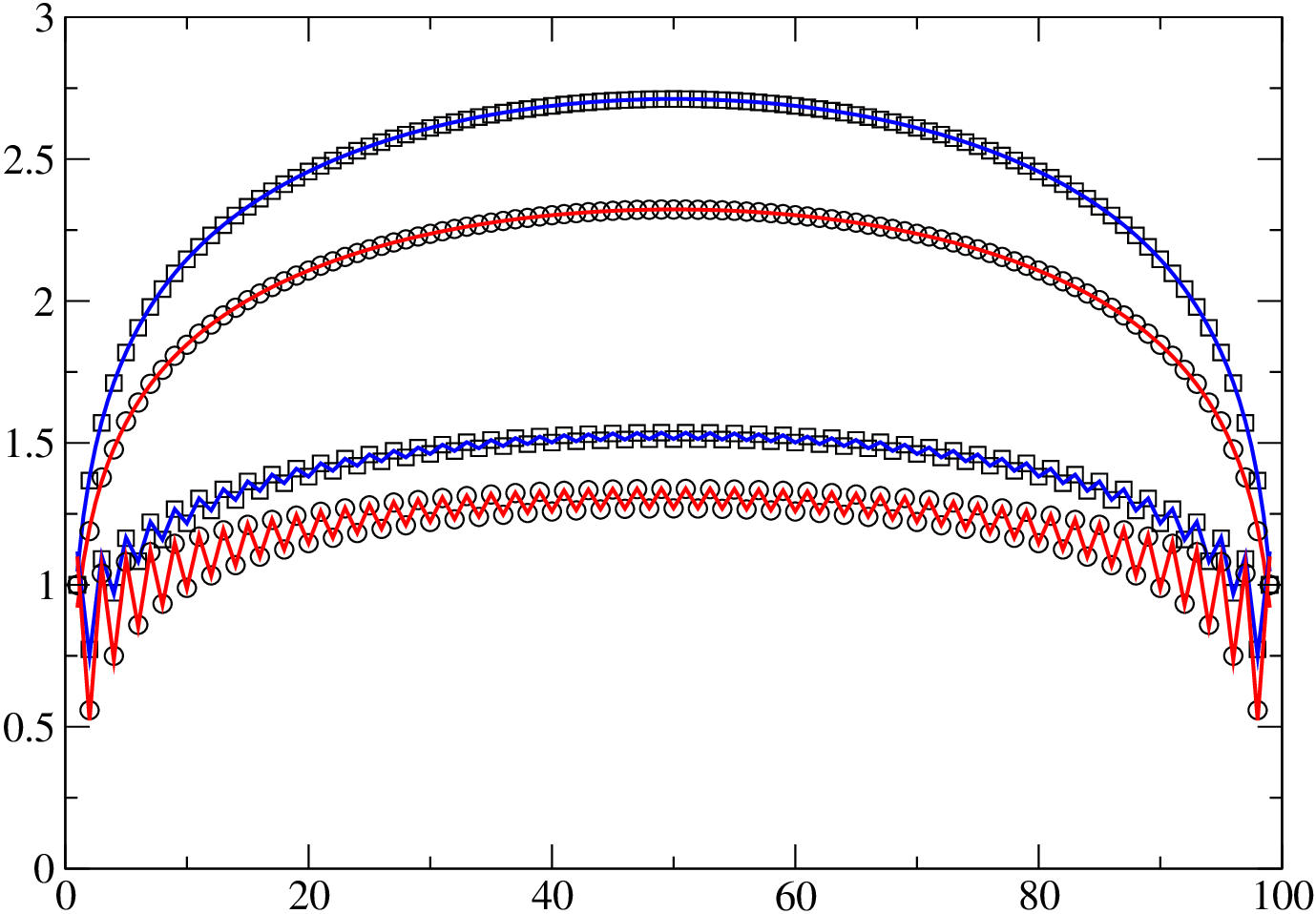}\\[-60pt]
\hspace{-2pt}
\includegraphics*[width=65pt]{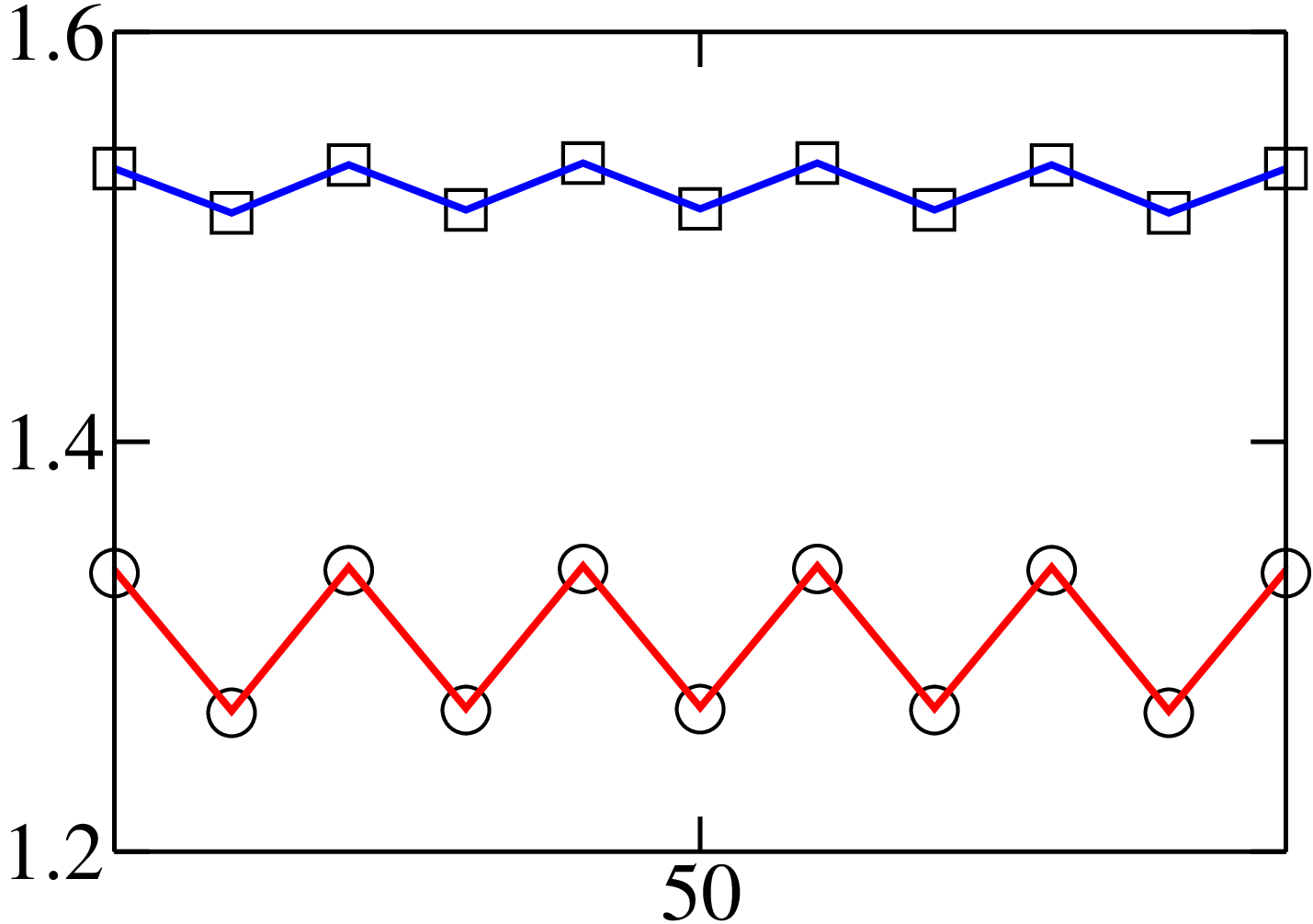}\\
\setlength{\unitlength}{1pt}
	\begin{picture}(0,0)(0,0)
		\put(77,125){\makebox(0,0){{\normalsize {PBC}}}}
		\put(77,35){\makebox(0,0){{\normalsize {OBC}}}}
		\put(2,-10){\makebox(0,0){{\normalsize $\ell$}}}
		\put(-123,40){\rotatebox{90}{{\normalsize $\mathcal{S}(\ell,L)$, $4\mathcal{F}(\ell,L)$}}}
	\end{picture}\\[5pt]
\caption{\label{fig:xx-100-S-F}(color online) Results from exact diagonalization for the entanglement entropy (squares) and spin fluctuations (circles) for a spin-1/2 XX chain with PBCs (upper symbols) and OBCs (lower symbols, inset shows zoom around the chain center), $L=100$. For PBCs solid lines are the analytical results in Eq.~(\ref{eq:SPBC}) and Eq.~(\ref{eq:FPBC}). For OBCs, solid lines are the analytical result (\ref{eq:FOBC}) for the fluctuations and a fit to Eq.~(\ref{eq:SOBC}) for the entropy with $c\simeq0.997$ and $s_1\simeq 1.050$ (the exact values are $c=1$ and $s_1\simeq 1.047$). The fluctuations are scaled by 4.}
\end{figure}
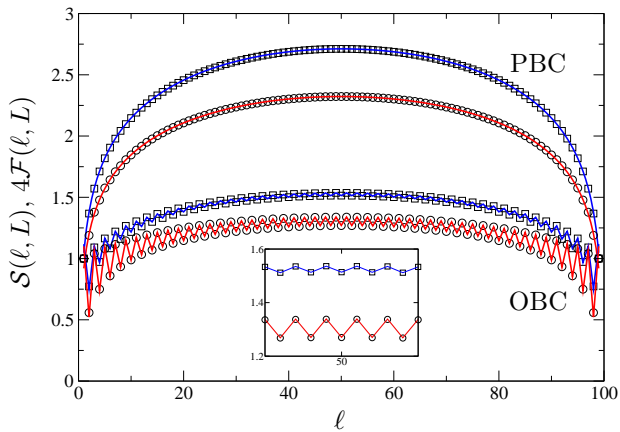

For $\ell\gg1$, an analytical result was obtained for the entanglement entropy for PBCs \cite{Sxx}:
\be
	\mathcal{S}_\text{XX}(\ell,L) = \frac{c}{3}\log_2 \ell+ s_1, \label{eq:SPBC}
\ee where $c=1$, $s_1\simeq1.047$, and we use $\log_2$ for the entropy. For easier comparison to numerical data we always work with the formula for finite $L$ which corresponds to the mapping $\ell\rightarrow(L/\pi)\sin(\pi\ell/L)$. Similarly, we find that the spin fluctuations for PBCs are given by
\be
	\pi^2\mathcal{F}_\text{XX}(\ell,L) = \ln \ell + f_1\label{eq:FPBC}
\ee plus $O(\ell^{-2})$ corrections, where $f_1=1+\gamma+\ln2$ and $\gamma$ is Euler's constant. This is consistent with $K=1$ in the corresponding LL description. Eq.~\eqref{eq:FPBC} was also derived in a different context \cite{eisler}. Fig.~\ref{fig:xx-100-S-F} compares the exact diagonalization result to the analytical results for both the entanglement entropy and spin fluctuations for PBCs; even for 100 sites the agreement is excellent. 

For OBCs the spin fluctuations are given by 
\begin{align}
	\mathcal{F}^\text{obc}_\text{XX}(\ell,L)&=\frac{1}{2}\mathcal{F}_\text{XX}(2\ell,L) \notag\\
	&\qquad+ \frac{1}{2\pi^2(2\ell)} - \frac{(-1)^\ell}{\pi^2(2\ell)}[\ln(2\ell) + \gamma+\ln2]\label{eq:FOBC}
\end{align}plus $O(\ell^{-2})$ corrections. The result is very similar to the oscillating form found in Ref.~\onlinecite{sorensen} for the entanglement entropy in the presence of a boundary, with an additional oscillating contribution $\propto(-1)^\ell (\ln \ell)/\ell$. As shown in Fig.~\ref{fig:xx-100-S-F}, the entanglement entropy is described well by
\be
	\mathcal{S}^\text{obc}_\text{XX}(\ell,L) = \frac{c}{6}\log_2(2\ell) + \frac{s_1}{2}
		+ a_1\frac{1}{(2\ell)}- a_2\frac{(-1)^\ell}{(2\ell)}\label{eq:SOBC}.
\ee

The XXZ model is solvable for all $\Delta$ by Bethe ansatz, but extracting the exact fluctuations is not practical. However, from LL theory we know that for $|\Delta|<1$ its asymptotic behavior (for PBCs) is \cite{giamarchi}
\be
	\mean{\hat{S}^z_{i+r}\hat{S}^z_i}-\mean{\hat{S}^z_{i+r}}\mean{\hat{S}^z_i} = -\frac{K}{2\pi^2}\frac{1}{r^2} + \frac{2A_2}{\pi^2}\frac{(-1)^r}{r^{2K}} + \ldots\label{eq:SSxxz},
\ee where only the leading terms are shown with $K=(1/2)[1-(\cos^{-1}\Delta)/\pi]^{-1}$ and $A_2$ a non-universal constant. From Eq.~(\ref{eq:SSxxz}) we find
\be
	\pi^2\mathcal{F}_\text{XXZ}(\ell) = K\ln \ell  +f_2 - A_2\frac{(-1)^\ell}{\ell^{2K}} \label{eq:Fxxz}
\ee plus $O(\ell^{-2})$ corrections, while the entanglement entropy is given by Eq.~(\ref{eq:SPBC}) with $c=1$ but a different constant. In the derivation of Eq.~(\ref{eq:Fxxz}) a term proportional to $\ell$ was suppressed, since it arises from the short-distance physics not taken into account by Eq.~(\ref{eq:SSxxz}) and we are guaranteed by Eq.~\eqref{eq:LL} that the leading term is $\propto\ln\ell$. It is interesting that the logarithmic divergence arises from the $1/r^2$ term in the  correlation function, which for $K<1$ is the \emph{subleading} contribution at large $r$. Thus the diverging fluctuations are due to short-distance correlations.

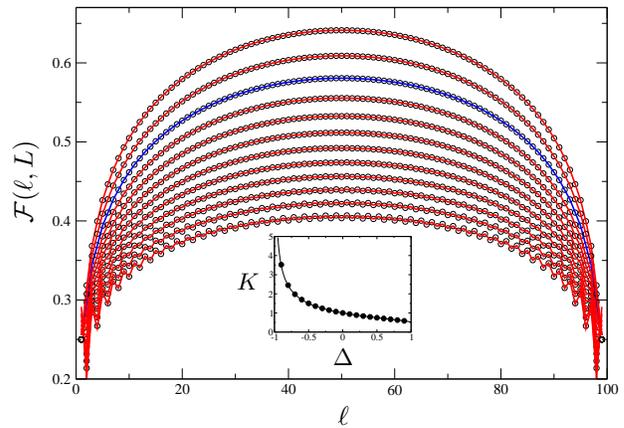
\begin{figure}[tp]
\includegraphics*[width=215pt]{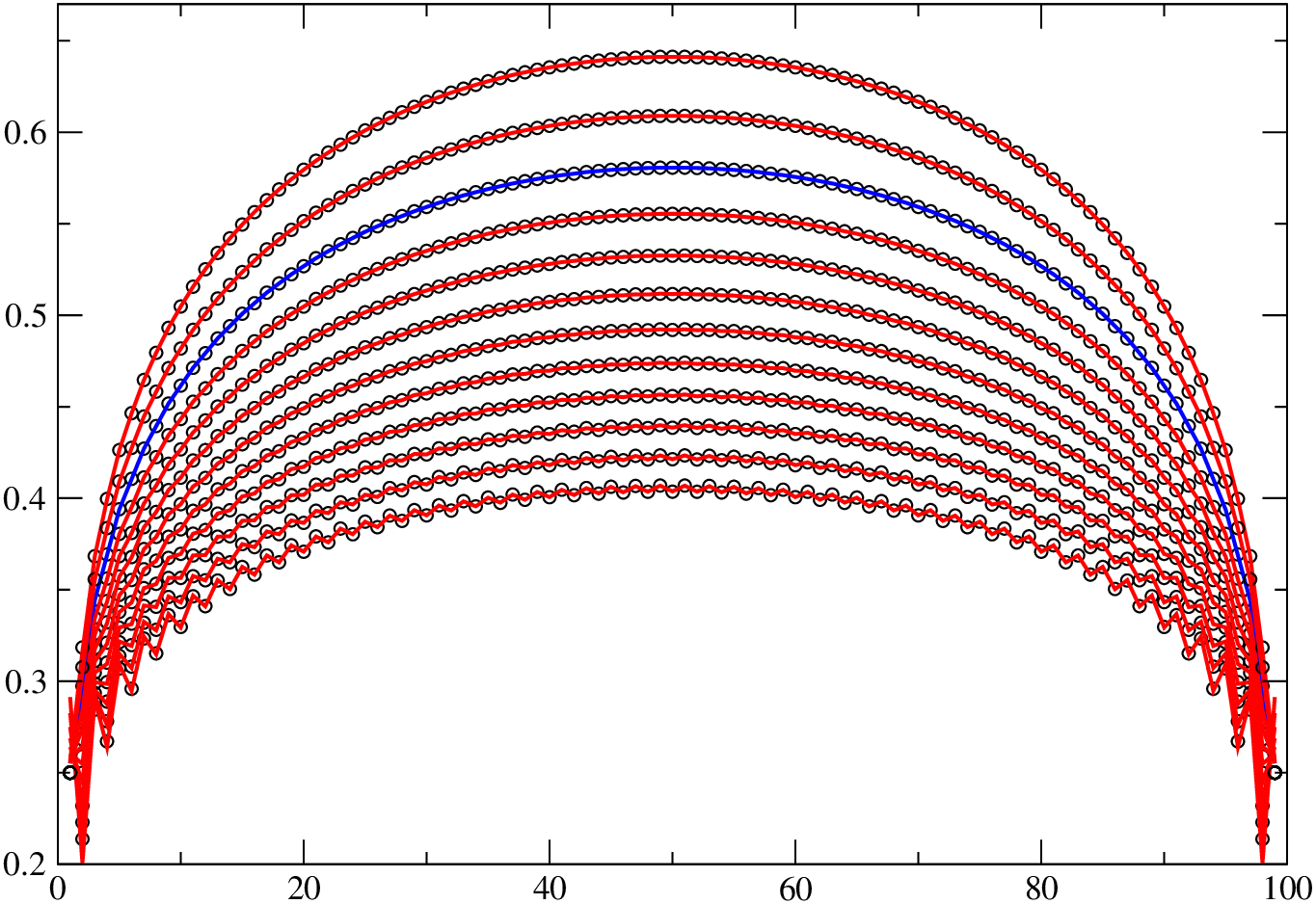}\\[-63pt]
\hspace{1pt}
\includegraphics*[width=55pt]{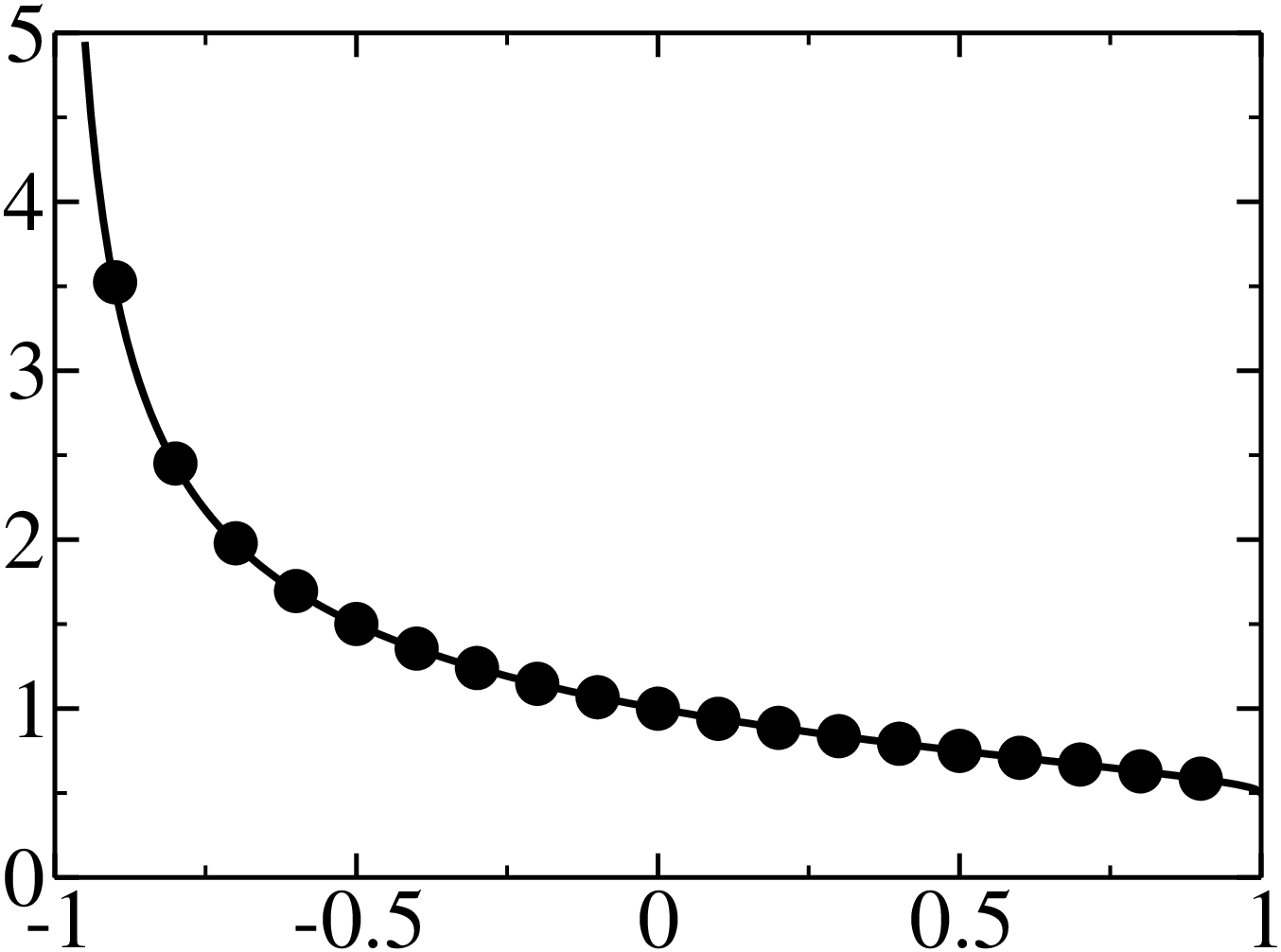}\\
\setlength{\unitlength}{1pt}
	\begin{picture}(0,0)(0,0)
		\put(3,4){\makebox(0,0){$\Delta$}}
		\put(-33,33){\makebox(0,0){$K$}}
		\put(3,-18){\makebox(0,0){{\normalsize $\ell$}}}
		\put(-122,55){\rotatebox{90}{{\normalsize $\mathcal{F}(\ell,L)$}}}
	\end{picture}\\[15pt]
\caption{\label{fig:xxz-pbc-100-F}(color online) DMRG results (circles) for the spin fluctuations in a spin-1/2 XXZ chain with PBCs, $L=100$. Only $\Delta=-0.2$ to 0.9 are shown (from top to bottom, in 0.1 increments). Solid lines are fits to Eq.~(\ref{eq:Fxxz}). (INSET) Fitted Luttinger parameter $K$ for $|\Delta|\leq0.9$ (dots), with the solid line showing the Bethe-ansatz solution. The first and last 10 sites were dropped for fitting purposes.}
\end{figure}

Since $K=1$ for the XX model the oscillating term can be neglected to $O(\ell^{-2})$ in agreement with our previous result (\ref{eq:FPBC}). The same is true for $\Delta<0$, because $K>1$. In contrast, for $\Delta>0$ the oscillations grow larger as we approach the Heisenberg point $\Delta=1$ where $K=1/2$. Interestingly, although the same oscillating terms are not present in the entanglement entropy itself, they were recently shown to be a feature of the R\'{e}nyi entropies \cite{renyi}. At exactly the Heisenberg point the otherwise irrelevant umklapp term $\cos(4\phi)$ becomes marginal, and the oscillating part of the spin-spin correlation function acquires a logarithmic correction $A_2'(-1)^r\sqrt{\ln r}/r$. In this case the corresponding term in the fluctuations is $\propto(-1)^\ell\sqrt{\ln\ell}/\ell$. More importantly for finite $L$ the Luttinger parameter $K$ is also renormalized. Fig.~\ref{fig:xxz-pbc-100-F} shows the fit of density-matrix renormalization group (DMRG) \cite{dmrg} data to Eq.~(\ref{eq:Fxxz}), with excellent agreement between the fitted $K$ and the Bethe ansatz solution for $|\Delta|\leq0.9$. 

\emph{Other spin chains.}---The isotropic Heisenberg model at $\Delta=1$ is also an example of the SU(2) Wess-Zumino-Witten (WZW) nonlinear $\sigma$-model with (integer) topological coupling constant $k$ \cite{witten}, which describes the low-energy physics of many other spin chains. Critical Heisenberg chains with half-integer spin belong to the $k=1$ universality class with central charge $c=1$ \cite{wzw}, for example, and we can deduce $\pi^2\mathcal{F}_\text{WZW}(x)\sim (k/2)\ln (x/a)$. An interesting case where we can compute the fluctuations analytically is the Haldane-Shastry (HS) model with $1/r^2$ interactions \cite{hsm}, which has exactly known spin-spin correlation function $\mean{\hat{S}^z_{i+r}\hat{S}^z_i} - \mean{\hat{S}^z_{i+r}}\mean{\hat{S}^z_i} = (-1)^r \text{Si}(\pi r)/(4\pi r)$ with $\text{Si}(x)=\int_0^x dt\ (\sin t)/t$. For large $\ell$ we find
\be
	\mathcal{F}_\text{HS}(\ell) = \frac{1}{2\pi^2}\ln\ell + f_3 - \frac{(-1)^\ell}{16\ell}\label{eq:HSM-f}
\ee plus $O(\ell^{-2})$ corrections, where $f_3$ is an integral whose value is $f_3\simeq 0.197$. This is consistent with the WZW fixed-point predictions, and in particular with the spin-1/2 Heisenberg chain \emph{without} the umklapp term. This also shows that the logarithmic scaling is not affected by long-range interactions.

A case where $k>1$ is the spin-$s$ Takhtajan-Babujan (TB) chain \cite{wzw}. Using DMRG we have checked the case $k=2s=2$ and hence $\pi^2\mathcal{F}_\text{TB}(x)\sim\ln (x/a)$, with corresponding central charge $c=3k/(2+k)=3/2$. This also explicitly confirms the result $g=\pi v\chi=s=k/2$ in Eq.~\eqref{eq:gfactor}. Another interesting example is the open-boundary Uimin-Sutherland model which is a critical spin-1 chain with SU(3) symmetry, because both the entanglement entropy and spin fluctuations exhibit oscillations with a period of 3 sites due to the higher symmetry.

\emph{Bose-Hubbard Model.}---The Hamiltonian for the one-dimensional Bose-Hubbard model is \cite{fisher}
\be
	H_\text{BH} = -t\sum_i (\hat{b}^\dag_i\hat{b}^{}_{i+1} + \text{h.c.}) + \frac{U}{2}\sum_i \hat{n}_i(\hat{n}_i-1)\label{eq:BH},
\ee where $\hat{b}_i$ is the bosonic annihilation operator on site $i$, $\hat{n}_i=\hat{b}^\dag_i\hat{b}^{}_i$, $t$ is the tunneling amplitude and $U$ is the on-site repulsion. The model describes interacting bosons on a lattice, and can be realized with cold atoms trapped in an optical lattice \cite{tonks}. As an example, we consider the case of half-filling since it approaches the XX model as $U\rightarrow\infty$. The low-energy physics of the model is that of a LL \cite{giamarchi}, so that the density fluctuations are given also by Eq.~(\ref{eq:Fxxz}). However, because $K\geq1$ the oscillations are absent for PBCs and very weak for OBCs. We have confirmed all of these results with DMRG and obtained curves similar to those shown in Fig.~\ref{fig:xx-100-S-F} for the XX model.

\emph{Gapped models.}---For gapped models, we expect that the fluctuations will obey an area law like entanglement entropy, although the ratio is no longer fixed by conformal arguments. This can be checked explicitly for the Affleck-Kennedy-Lieb-Tasaki model \cite{aklt}, for example, where analytical results are available but also the valence bond picture makes the relation intuitive \cite{disorder}.

\emph{Conclusion.}---From the theory of quantum critical phenomena one expects an intimate relation between entanglement and fluctuations. In one dimension this expectation is borne out by Eq.~(\ref{eq:SFratio}): like entanglement entropy, the number fluctuations scale logarithmically in critical models described by a CFT with a globally conserved charge. Moreover, our detailed investigation of the effects of boundaries and subleading corrections for several important models suggests that studying fluctuations is a powerful approach to understanding the scaling of entanglement entropy more generally. This has clear advantages: First, fluctuations are accessible in experiments, perhaps most easily for cold atoms in an optical lattice. Second, from the computational point of view they are easier to calculate, both analytically and numerically, than entanglement entropy. In particular, the close connection between entanglement entropy and fluctuations suggests that investigations of the latter with quantum Monte Carlo for $d>1$ will provide clues to such important questions as whether the $L^{d-1}\ln L$ scaling of both $\mathcal{S}$ and $\mathcal{F}$ in free fermions \cite{klich} holds more generally for interacting fermions \cite{swingle}.

We thank Steve Girvin, Leonid Glazman, and Nick Read for valuable discussions, and Peter Schmitteckert for use of his DMRG code. This work was supported by NSF Grant No. DMR-0803200 and by the Yale Center for Quantum Information Physics (DMR-0653377). SR acknowledges support from the Deutsche Forschungsgemeinschaft under Grant No. RA 1949/1-1.

\vspace{-10pt}

\bibliographystyle{h-physrev}

\end{document}